\documentclass[iop, apj]{emulateapj}  
\usepackage{natbib}
\bibpunct{(}{)}{;}{a}{}{,} % to follow the A&A style

\usepackage{graphicx}
\usepackage{amsmath}
\usepackage{txfonts}

\begin{document}

   \title{Alfv\'en Wave Heating of the Solar Chromosphere: 1.5D models}

   \author{T.D.~Arber$^1$, C.S.~Brady$^1$, S.~Shelyag$^2$}

   \affil{$^1$Centre for Fusion, Space and Astrophysics
	University of Warwick, Coventry, CV4 7AL, UK}
   \affil{$^2$School of Mathematical Sciences, Monash University, Clayton, 3800, Australia}

\begin{abstract}
Physical processes which may lead to solar chromospheric heating are analyzed using high-resolution 1.5D non-ideal MHD modelling. 
We demonstrate that it is possible to heat the chromospheric plasma by direct resistive dissipation of 
high-frequency Alfv\'en waves through Pedersen resistivity. However this is unlikely to be sufficient to balance radiative and conductive
losses unless unrealistic field strengths or photospheric velocities are used.
The precise heating profile is determined by the input driving spectrum
since in 1.5D there is no possibility of Alfv\'en wave turbulence. The inclusion of the Hall term does not affect the heating rates.
If plasma compressibility is taken into account, shocks are produced 
through the ponderomotive coupling of Alfv\'en waves to slow modes and shock heating dominates the resistive dissipation. 
In 1.5D shock coalescence amplifies the effects of shocks and for compressible simulations with realistic
driver spectra the heating rate exceeds that required to match radiative and conductive losses. Thus while the heating rates for these 1.5D simulations are an overestimate they do show that ponderomotive coupling of Alfv\'en waves
to sound waves is more important in chromospheric heating than Pedersen dissipation through ion-neutral collisions.

\end{abstract}

\keywords{Sun: atmosphere, Sun: chromosphere}
%
%________________________________________________________________

\section{Introduction}
Alfv\'en waves have been proposed as an energy delivery mechanism for solar atmospheric processes ranging from heating of the chromosphere 
\citep{Osterbrock1961,Ballegooijen2011} or corona \citep{Ballegooijen2011,Kudoh1999} to the acceleration of the solar wind \citep{DePontieu2007}. 
Both observations \citep{DePontieu2007} and numerical models \citep[e.g.][]{Shelyag2012, Shelyag2013} show that there is enough Alfv\'en wave energy 
generated in the convection zone to heat the chromosphere, but 
the mechanisms by which this wave energy is thermalized in the chromosphere, if indeed it is, remain unclear. If Alfv\'en waves are a dominant source 
of quiet chromospheric heating the observed input flux of $\sim 10^{7} \mathrm{\:erg\:cm^{-2}\:s^{-1}}$ must be converted into a 
$\sim 0.1 \mathrm{\:erg\:cm^{-3}\:s^{-1}}$ local heating rate to balance the energy loss in the chromosphere. Estimates place the maximum
heating rate required as $\sim 0.1 \mathrm{\:erg\:cm^{-3}\:s^{-1}}$ dropping to $\sim 10^{-3} \mathrm{\:erg\:cm^{-3}\:s^{-1}}$ in the upper
chromosphere \cite{Ulmschneider1974,Avrett1981}.

The possibility that Alfv\'en waves may heat the solar atmosphere and drive the solar wind has been studied numerically with many variations.
\citet{Hollweg1987,Hollweg1981,Hollweg1982,Hollweg1992} extensively studied the 1.5D problem for circularly polarized Alfv\'en waves in expanding 
flux tubes and 
showed the importance of coupling to slow modes, fast modes and shocks. These papers demonstrated that it was certainly possible for 
Alfv\'en waves to heat
the solar atmosphere and drive the solar wind. Later work has extended this to white noise spectrum photospheric drivers in both 1.5D \citep{Kudoh1999}
and 2.5D \citep{Matsumoto2012}. The main focus of these papers has been on coronal heating and driving the solar wind. Thus, while they contained model
atmospheres with chromospheres and transition regions, these were not modelled with very high resolution, nor were detailed heating profiles produced 
specifically for the chromosphere. Despite this, all papers listed above did show that a heating rate broadly consistent with the requirements of
both the chromosphere and corona was possible for photospheric motions with speeds of order $\sim 1$ km s$^{-1}$ driving Alfv\'enic perturbations.

\citet{Matsumoto2014} extended their original analysis \citep{Matsumoto2012} and improved the diagnostics for shock heating and concluded that
shock heating was the dominant heating mechanism in the chromosphere. These simulations did not explicitly include resistivity or assess the 
sensitivity of these results to the chosen driver spectrum. Nonetheless they present a compelling case 
for shock heating of the chromosphere originating from a broad spectrum photospheric, Alfv\'enic driver. The results presented below represent
an extension of this line of work, albeit only in one spatial dimension, to assess the importance of explicit resistive dissipation through 
electron-ion, electron-neutral and ion-neutral collisions. Also studied is the sensitivity of the results to the chosen driver spectrum and a high-resolution focus on 
just the chromosphere.

Turbulence has long been proposed as a possible mechanism by which the input driver wave energy could be dissipated \cite{Hollweg1986}. 
Direct dissipation of driver Alfv\'en waves through resistive effects is too slow to explain the required heating rate. However, if a turbulent
cascade exists which can transfer the driver energy to higher wavenumbers where dissipation does occur there is sufficient wave energy in the chromosphere
to balance radiative losses. This theory does however leave unanswered the question of whether a turbulent cascade to dissipation scales does actually exist
in the chromosphere.
Reduced MHD models have recently concluded that Alfv\'en waves may drive a turbulent
cascade, through reflection off the transition region, which could heat the chromosphere \citep{Ballegooijen2011}. The
implementation of reduced MHD in \citet{Ballegooijen2011} ignores compressive effects so does not permit coupling of
Alfv\'en waves to slow modes or the formation of shocks.

Studies of the resistive dissipation of Alfv\'en waves by collisions between ions and neutral atoms, begun with \citet{Piddington1956},
have recently begun to be re-investigated and extended
 \citep{dePontieu2001,Leake2005,Goodman2011,Tu2013}. The temperature variation through the chromosphere means that it contains both regions where the 
hydrogen is near fully ionized and regions where only heavy ions with a low first ionization potential are ionized \citep{Fontenla1993, Vernazza1981, 
Avrett2008}. Under low-ionization conditions the ion and neutral gas fluids partly decouple, and energy dissipation due to 
ion-neutral collisions is orders
of magnitude more important than the resistivity resulting from electron-ion Coulomb collisions  
\citep{khodachenko2004, Khomenko2012}. Previous work in this area has shown efficient heating of both the upper chromosphere 
\citep{dePontieu2001,Leake2005} and the lower chromosphere \citep{Goodman2011,Song2011,Tu2013} due to ion-neutral collisions. 

Numerical simulations in this area are complicated by the 
simultaneous requirements to resolve short driver wavelengths and a high Alfv\'en speed in the upper chromosphere. \citet{Tu2013} also include the 
Hall term in chromospheric heating models further reducing the timestep in the simulations. The computational effort that is required to simultaneously 
resolve all of these length and time scales means that 1.5D modeling is often used to model Alfv\'en wave propagation and heating in the chromosphere
\citep{Leake2005,Tu2013}. These have shown that Alfv\'en waves with periods much less than 10 seconds are effectively damped by Pedersen resistivity
\citep{dePontieu2001,Leake2005} in the chromosphere, and for incompressible simulations with
a Kolmogorov driving spectrum the heating is primarily in the lower chromosphere and 
sufficient to balance radiative losses there \citep{Tu2013}.

There is therefore compelling observational and theoretical evidence that a significant fraction of the heating requirements of the upper chromosphere, 
and possibly the quiet corona, could be accounted for through the dissipation of Alfv\'en wave energy, or MHD waves in general in geometries which 
make pure Alfv\'en waves unlikely. This paper extends the earlier works by concentrating specifically on the chromosphere and varying the spectrum of 
driven Alfv\'en waves, solving the full compressible MHD equations and including partially ionized plasmas. Section \ref{sec:numtech} 
covers the assumptions made, the numerical method and driver spectra used in the simulations. Since much previous work on turbulent heating 
\citep{Tu2013,Ballegooijen2011} assumed an incompressible plasma incompressible results are presented in section \ref{sec:incompressible}. 
Full compressible
simulations results are presented in section \ref{sec:compressible} and conclusions which can be drawn from these 1.5D simulations are
in section \ref{sec:conclusions}.

\section{MODEL EQUATIONS}
\label{sec:numtech}
This paper presents the results of numerical simulations performed using the Hall-MHD code LARE \citep{Arber2001}. 
This code solves the ideal MHD 
equations explicitly using the Lagrangian remap approach and includes the resistive and Hall terms using explicit subcycling. 
The complete set of MHD equations relevant in this case are
 
\begin{eqnarray}
	\frac{D \rho}{Dt} &=& -\rho\nabla.\mathbf{v} \\
	\rho \frac{D \mathbf{v}}{Dt}& =& \mathbf{j} \times \mathbf{B} +\rho \mathbf{g}- \nabla P + \mathbf{F}_{shock}
	\label{eq:momentum}\\
	\frac{\partial \mathbf{B}}{\partial t} &= &-\nabla \times \mathbf{E}\\
	\frac{D \epsilon}{Dt} &=& -\frac{P}{\rho} \nabla . \mathbf{v} + \frac{H_{\mathrm{visc}}}{\rho} + \frac{H_{\mathrm{Ohmic}}}{\rho}\label{eq:energy}\\ 
	\mathbf{j}&=&\frac{1}{\mu_0}\nabla \times \mathbf{B}\\
	\mathbf{E} &=& - \mathbf{v} \times \mathbf{B} + {\eta}\mathbf{j}_\parallel + \eta_p\mathbf{j _{\perp}} + \frac{1}{e n_e}\mathbf{j} 
	\times \mathbf{B} \label{eq:ohm}
\end{eqnarray}

where $\rho$ is the mass density, $P$ is the ideal gas pressure, $\mathbf{B}$ is the 
magnetic field, $\mathbf{v}$ is the fluid velocity, $\mathbf{E}$ is the electric field,  $\epsilon$ is the specific internal energy density, 
$n_e$ is the electron number density, $e$ is the proton charge and $(\mathbf{j_\parallel},\mathbf{j_\perp})$ are the current densities 
parallel/perpendicular to the local 
magnetic field. These equations are in written in Lagrangian form where
\begin{equation*}
\frac{D }{Dt}=\frac{\partial }{\partial t}+\mathbf{v}.\nabla
\end{equation*}
is the usual advective derivative. Shock viscosity is added to the momentum equation through $\mathbf{F}_{shock}$ which has a 
functional form which vanishes for smooth flows in the limit of increasing resolution but remains finite at discontinuities. Thus 
the shock jump conditions are satisfied through an appropriate shock viscosity as in \citet{Caramana1998, Arber2001}. In this
formulation shock heating can therefore be monitored though viscous heating. Hence viscous heating and shock heating are used as 
synonyms throughout the rest of this paper.

Missing from Equation~(\ref{eq:energy}) are thermal conduction, radiative losses and all non-local radiation transport. To prevent the 
atmosphere heating up without bounds due to the viscous dissipation of shocks ($H_{visc}$) and Ohmic heating 
($H_{Ohmic}$), both $H_{visc}$ and $H_{Ohmic}$ are calculated in simulations as a diagnostic but only $H_{visc}$ is added to the energy update. 
If $H_{visc}=0$, as in incompressible simulations, ignoring $H_{Ohmic}$ assumes that the background atmosphere is for a slowly
varying chromosphere where heating terms are approximately balanced by losses. Shock heating must be included in compressible simulations
otherwise the shocks themselves are not treated correctly. This leads to a heating of the chromosphere. Despite this the ionization state is held fixed
at its initial equilibrium value. Since shock heating, which is only weakly sensitive to the ionization state, dominates in these simulations this
will not change the heating rates by more than a factor $\sim 2$. 

The resistive MHD Ohm's law for a fully ionized plasma is $\mathbf{E} = - \mathbf{v} \times \mathbf{B} + 
{\eta_\parallel}\mathbf{j}_\parallel + \eta_\perp \mathbf{j _{\perp}}$ 
\citep{braginskii} where 
\begin{equation*}
\eta_\perp=\frac{m_e \nu_{ei}}{n_e e^2}
\end{equation*}
is the classical electron-ion resistivity resulting from Coulomb scattering and $\nu_{ei}=3.7\times n_i \log(\Lambda) / T^{3/2}$ 
is the electron-ion collision frequency for ion number density $n_i$ and Coulomb logarithm $\log(\Lambda)$. $\eta_\parallel = 0.51 \eta_\perp$ 
and the resistivity is anisotropic. Normally
within solar and space physics, where the Lundquist number is high, this is simplified to 
$\mathbf{E} = - \mathbf{v} \times \mathbf{B} + \eta_\perp \mathbf{j}$ with an isotropic resistivity. This approximations is adopted in this paper
so Ohmic dissipation resulting from parallel currents is likely to be over estimated by a factor $\simeq 2$. In the 1.5D 
simulations in this paper there can be no current along the background magnetic field so the leading order Ohmic dissipation from 
electron-ion collisions is for perpendicular currents.

In Equation~(\ref{eq:ohm}), $\eta$ is the classical  resistivity resulting from electron collisions with ions and neutrals
and $\eta_p$ the Pedersen resistivity \citep{cowling,Leake2005}. Explicitly the resistivities used in this paper are:
\begin{equation*}
\eta=\frac{m_e \nu_e}{n_e e^2}
\end{equation*}
where $\nu_e$ is the  electron collision frequency including collisions with ion and neutrals. The Pedersen resistivity is then given by
\begin{equation}
\label{Pedersen}
\eta_p=\eta+\frac{\xi_n^2 B^2}{(1-\xi_n)}\frac{1}{\rho \nu_{in}}
\end{equation}
where $\xi_n=m_i n_n / \rho$ is the neutral fraction with $n_n$  the number density of neutrals. The ion-neutral collision frequency is given by
\begin{equation*}
\nu_{in}=n_n \sqrt{\frac{16k_B T}{\pi m_i}}\Sigma_{in}
\end{equation*}
here $\Sigma_{in}=5\times 10^{-19} \mbox{m}^2$ is the cross-section for ion-neutral collisions for hydrogen from \cite{Osterbrock1961,DePontieu1998}.
For an effective ion mass of 1.2 proton masses $\nu_{en}\simeq 5 \nu_{in}$.
The pressure gradient term $-\nabla P_e$ is ignored in equation \ref{eq:ohm} as this can only
generate a parallel electric field in 1.5D simulations which have no effect on the dynamic equations. 

In the above descriptions of resistivities we
have adopted the nomenclature used in \citep{Leake2014}. Following \citep{Leake2014} we also use Cowling resistivity ($\eta_c$), defined as
$\eta_c=\eta_p - \eta$, to refer to that component of perpendicular resistivity which arises purely due to ion-neutral collisions.
It is important to note that the Pedersen resistivity is only the inverse of the Pedersen 
conductivity, more commonly used in ionospheric physics, if the Hall term is zero. 

The inclusion of neutrals in a single-fluid model, as described above, 
is valid in the same limit as resistive MHD, i.e. timescales larger than the gyro-period, collision times and the plasma period and 
length-scales larger than the ion gyroradius and collisional mean-free-path. The only additional constraint, compared to single fluid MHD, is 
therefore that equations
(1-6) are only valid for times larger than the ion-neutral collision time. In the chromosphere this can be as low as a millisecond so the model is valid
up to frequencies of $\simeq 10^3$ Hz.

In all simulations the $x$ axis is ignorable, the $y$ axis extends vertically up from the solar surface starting at the $\tau=1$ optically 
thick surface, and the $z$ axis is also ignorable, while velocities in the $z$ direction are allowed and associated with Alfv\'enic perturbations. 
The upper boundary condition is kept open by the use of both Riemann characteristic open boundaries and a damping region where the simulation 
velocity is smoothly reduced as it approaches the boundary. The measured reflection of the upper boundary is less than 0.01\%. In some sections 
LARE is converted into a 1.5D incompressible code by setting the forces parallel to the $y$ axis to zero. This sets $\mathrm{v}_y=0$, which 
in 1.5D is equivalent to the  incompressible 
condition $\nabla \cdot \mathbf{v}=0$.

For the simulations presented in this paper, a model atmosphere based on the semi-empirical C7 model of Avrett \& Loeser \citep{Avrett2008} is 
used for the initial temperature profile. This is then integrated iteratively to provide a density profile which is in hydrostatic equilibrium. 
On each iteration the ionization state of the plasma is calculated under the assumption of local thermal equilibrium using a two level Athay 
potential model \citep{Leake2006} until a converged density is obtained. The background magnetic field $B_y$ is assumed to be a uniform 
vertical 50G. This is chosen as a representative value averaged along an inter-granular field line. 

\begin{figure}
\includegraphics[scale=0.5]{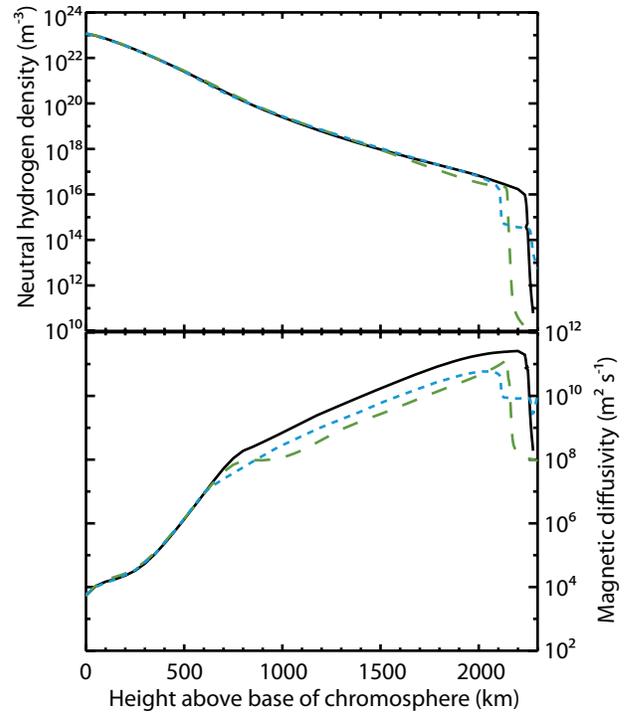}
\caption{a) Calculated neutral hydrogen density as a function of height above the base of the photosphere for different atmospheric models. 
b) Magnetic diffusivity ($\eta_p/\mu_0$) as a function of height above the base of the photosphere for different atmospheric models. 
Black line corresponds to FAL A \citep{Fontenla1993},  green long dashed line - to Avrett and Loeser C7 \citep{Avrett2008}, and blue short 
dashed line - to VAL C \citep{Vernazza1981} model.}
\label{fig:eta}
\end{figure}

\begin{figure}
\includegraphics[scale=0.4]{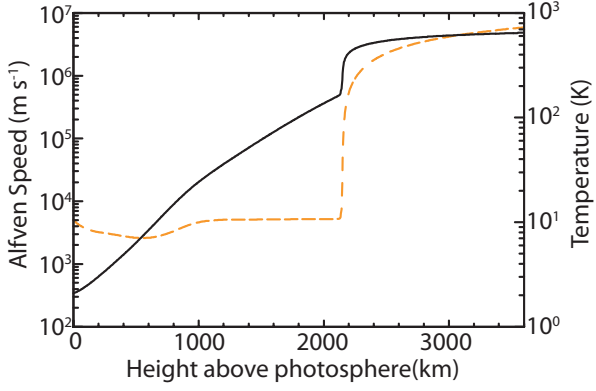}
\caption{Temperature profile (dashed line) and Alfv\'en speed (solid line) for a 50 G background field and the Avrett and Loeser C7 
atmosphere model.}
\label{fig:c_a}
\end{figure}

Figure~\ref{fig:eta} plots neutral density and Pedersen diffusivity 
for the Avrett and Loeser C7 atmosphere model in this paper. For comparison, Figure~\ref{fig:eta} also shows the same plots for
the two other atmospheric models: the VAL III C model of \citet{Vernazza1981} and the FAL A model of \citet{Fontenla1993}. It is found 
that the density of neutral hydrogen predicted by the Athay potential model and the associated Pedersen resistivity are 
similar in all cases although the FAL A model does predict 5 times greater Pedersen resistivity in the upper chromosphere. 
Figure~\ref{fig:c_a} shows the 
temperature and Alfv\'en speed profiles for the Avrett and Loeser C7 atmosphere model with an imposed background 50 G field.

\subsection{Boundary driving and Poynting flux}
Alfv\'en waves are introduced into the domain by driving the bottom boundary of the simulation. The velocity field is prescribed by 
Equation~(\ref{eq:driver}) 
to give a kinetic energy spectrum consisting of two regions - the low frequency region, where the power in a mode increases with 
frequency, and a Kolmogorov region at higher frequency, where power drops as $k^{-5/3}$:

\begin{equation}
	v_z=A (\sum \limits_{i=0} \limits^{N_1} \omega_i^\frac{5}{6} \sin{(\omega_i t + \phi_i)} + \sum \limits_{j=0} 
	\limits^{N_2} \omega_j^{-\frac{5}{6}} \sin{(\omega_j t + \phi_j)})
	\label{eq:driver}
\end{equation}
$\omega_i$ is the frequency of mode $i$ in the low frequency region and in this paper runs from 0.001Hz to 0.01Hz, $\omega_j$ is the frequency 
of mode $j$ in the high frequency region and in this paper runs from 0.01Hz to an arbitrary upper cutoff frequency which ranges between 0.1Hz 
and 10Hz. $\omega_i$ and $\omega_j$ are both linearly spaced in frequency. $v_z$ is the velocity component out of the plane of the simulation 
which is associated with an Alfv\'en wave. $N$ is the number of harmonics used to produce the driver spectrum and is a large 
enough number to ensure that the spectrum is reproduced smoothly and that further increase in $N$ does not lead to changes in the heating rate 
greater than 1\%. $N_1+N_2$ is set to 10,000 for simulations with a 10Hz upper cutoff frequency, 1000 for simulations with a 1Hz upper cutoff 
frequency and 100 for simulations with a 0.1 Hz upper cutoff frequency. Convergence testing using 100,000 driver elements for a simulation 
with a 10Hz upper cutoff shows that the result is not sensitive to this parameter so long as the total Poynting flux through the photospheric 
lower boundary is 
kept constant. $\phi_{i,j}$ is the phase for spectral component $i$ or $j$ and is selected randomly for each mode. 

\begin{figure}
\includegraphics[scale=0.5]{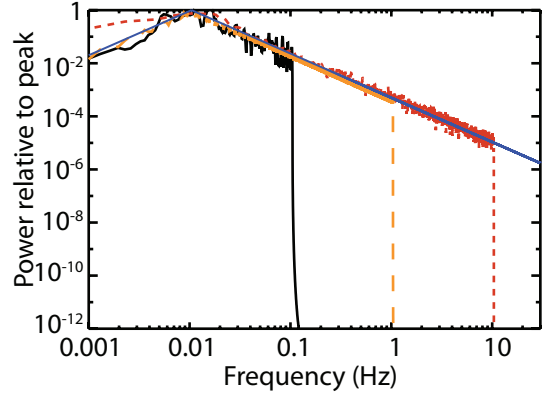}
\caption{Driver spectra with different upper cutoff frequencies. 10Hz (Red short dashed line), 1Hz (Orange long dashed line) and 0.1Hz 
(Black solid line) upper cutoff are shown with overplotted blue solid lines showing the envelope from the two terms in Equation~(\ref{eq:driver}).}
\label{fig:driver}
\end{figure}

The amplitude $A$ is selected to give a total energy flux at the bottom boundary of $10^{7} \mathrm{erg \:cm^{-2} \:s^{-1}}$. This corresponds 
to a velocity field with an RMS value of $415 \mathrm{m \:s^{-1}}$ at the $\tau=1$ optically thick photospheric surface, comparable with observations of 
photospheric transverse velocities (see for example \citet{Chae2001} or \citet{Nindos2002}).  The spectra of the used drivers, for differing 
cut-off frequencies, are show in Figure~\ref{fig:driver}. This form of the driver is based on that used by \citet{Tu2013}, although they
specified a net average Poynting flux of $2\times 10^{7} \mathrm{erg \:cm^{-2} \:s^{-1}}$. There is no observational evidence for such a
driver spectrum. Instead this has been chosen so that the r.m.s. velocity matches observations along with the total Poynting flux, which
constrains the magnetic field once photospheric densities and velocities are defined. Equation \ref{eq:driver} therefore has observationally
constrained average velocity and integrated flux. There is no direct evidence for high frequency Alfv\'en waves thus the choice of a
Kolmogorov spectrum is motivated purely by the observation that the photospheric motion is turbulent and hence some scale-free
power-law dependence is expected. The results are insensitive to the choice of spectrum in the rising part of the spectrum, 
which could have been chosen to be flat, other than its effect on the amplitude limiting to specify the net Poynting flux.

\begin{figure}
\includegraphics[scale=0.5]{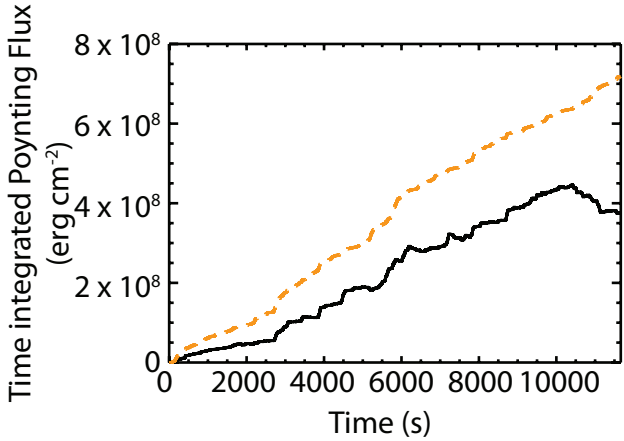}
\caption{Time-integrated Poynting flux through the lower boundary of a simulation of Alfv\'en waves propagating into the solar chromospheric cavity. 
The black solid line is calculated using Equation~(\ref{eq:fullpoynt}), orange dashed line using Equation~(\ref{eq:approxpoynt}).}
\label{fig:poynting_variants}
\end{figure}

The net Poynting flux from the specified velocity driving on the photosphere needs to be calculated from the simulations. 
In the single fluid MHD model this can be rewritten as 

\begin{equation}
	%\mathbf{S}(\mathbf{r},t)=\frac{1}{\mu_0}\mathbf{B}(\mathbf{r},t)\times(\mathbf{v}(\mathbf{r},t)\times\mathbf{B}(\mathbf{r},t))
\mathbf{S}(\mathbf{r},t)=\mu_0^{-1}\mathbf{B}(\mathbf{r},t)\times(\mathbf{v}(\mathbf{r},t)\times\mathbf{B}(\mathbf{r},t))	
\label{eq:fullpoynt}
\end{equation}
where $\mathbf{v}$ is the fluid centre of mass velocity, and $\mathbf{B}$ is the magnetic field. The commonly used approximation 
\begin{equation}
	S =  <v_z^2> c_A \rho
\label{eq:approxpoynt}
\end{equation}
where $v_z$ is the velocity amplitude perpendicular to the magnetic field and $c_A$ is the local Alfv\'en speed, fails for the
photospheric driver as this is a linear approximation for waves propagating only in one direction. At the photosphere, where there are 
reflected waves from the transition region, this approximation is not valid.
This is demonstrated in Figure~\ref{fig:poynting_variants}, where the time-integrated 
Poynting vector calculated using both Equations~(\ref{eq:approxpoynt}) and (\ref{eq:fullpoynt}) through the lower boundary of a 
simulation of the solar chromosphere with no dissipation mechanisms present are shown. These agree up until the first driven Alfv\'en
waves from the driver have returned after partial reflection from the transition regions, around 160 seconds, at which point they diverge.
Equation~(\ref{eq:approxpoynt}) overestimates the total Poynting flux through the boundary by a factor of up to 2. This means that 
estimates of Poynting flux obtained using 
Equation~(\ref{eq:approxpoynt}), whether from simulation or observation, are almost certain to be overestimated. In this paper the Poynting 
flux is always calculated using the form in Equation~(\ref{eq:fullpoynt}).\\

\section{Incompressible model}
\label{sec:incompressible}

Using the 1.5D setup as previously described and assuming that the plasma is incompressible ($\nabla \cdot \mathbf{v} =0, v_y=\mathrm{const}$), 
the MHD equations (\ref{eq:momentum})-(\ref{eq:ohm}) simplify to

\begin{eqnarray}
	\rho \frac{\partial v_z}{\partial t}& =& \frac{1}{\mu_0}B_y \frac{\partial B_z}{\partial y}, \\
	\frac{\partial B_z}{\partial t} &= & B_y \frac{\partial v_z}{\partial y}.
\end{eqnarray}
Normalizing to the constant vertical field and the local Alfv\'en speed such that $B_z=B_y b$ and $v_z=c_A v'$ it is possible to rewrite 
these equations in 
terms of the Els{\"a}sser variables $z^\pm = v \pm b $, where $z^-$ and $z^+$ correspond to upwards and downwards propagating Alfv\'{e}n 
waves respectively. After dropping the primes for normalized variable equations (11-12) are,

\begin{equation}
	\frac{\partial z^\pm}{\partial t} \mp c_A \frac{\partial z^\pm}{\partial y}=
	\pm \frac{(z^+ + z^-)}{2} \frac{\partial c_A}{\partial y}.
\end{equation}

There is no non-linear coupling between upward 
propagating and downward propagating waves which would lead to a cascade, i.e. no non-linear interaction between $z^+$ and $z^-$. 
Alfv\'en turbulence is suppressed in 1.5D by the assumption of incompressibility. 
Thus energy introduced by the driver in a given frequency will stay in that frequency until it is either dissipated or leaves the computational domain. 
LARE is converted to a 1.5D incompressible code as described 
in Section~\ref{sec:numtech} and used to simulate an Alfv\'en wave spectrum given by Equation~(\ref{eq:driver}) propagating into the Avrett \& Loeser 
C7 model atmosphere. 4096 grid points are used in the $y$ direction, and a length of 6000~km above the 
visible solar surface is simulated for all results, even those for which results are only plotted for the chromosphere.
At heights above 
4500km the velocity damping region is applied. Convergence is tested with a simulation of 8192 grid points, resulting in a change in the solution 
of $<1\%$ as measured by the heating rate. 

\begin{figure}
\includegraphics[scale=0.5]{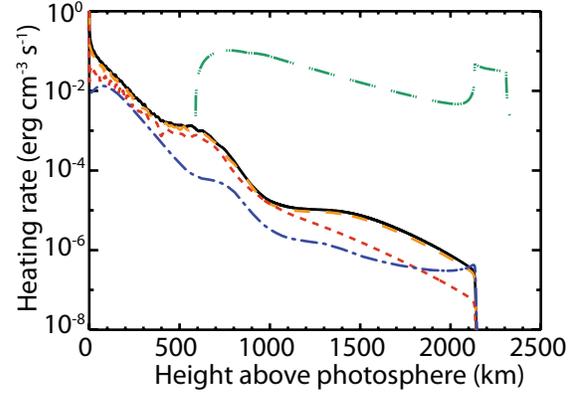}
\caption{Heating rate as a function of height for incompressible simulations of Alfv\'en waves driven into a C7 model atmosphere. 
The upper cutoff frequency is 0.1Hz in the simulation shown by the red short dashed line, 1Hz by the orange long dashed line, and 10Hz by the black solid line. 
The blue line is the analytic estimate from Equation~(\ref{eq:heating_analytic}). The green dot-dot-dash line is the estimated heating requirement
for the quite chromosphere taken from \citep{Avrett1981}.}
\label{fig:heating_height}
\end{figure}

Simulating an Alfv\'{e}n wave spectrum given by Equation~(\ref{eq:driver})  with a high frequency cutoff of 0.1Hz, 1Hz or 10Hz propagating up 
into the C7 atmosphere (Figure~\ref{fig:heating_height}), it is found that heating is highest in the photosphere with a local maximum at the 
temperature minimum. This is comparable to the results in 
\citet{Goodman2011} and \citet{Tu2013} 
who used a similar model to that presented here. Since resistive 
dissipation is more effective for shorter wavelengths and there is no turbulent cascade, the choice of upper cutoff frequency for the 
driver would be expected to affect the efficiency of the dissipation. This is observed in Figure~\ref{fig:heating_height} where greater 
heating at all heights is observed if a higher upper cutoff frequency in the driver spectrum is chosen. The simulations produce heating 
rates comparable to those in \citet{Tu2013}, with or without the Hall term included in Ohm's law. The Hall term is found to have negligible 
effect ($<0.01\%$) on the heating rate in all simulations. Note that the three lines in figure \ref{fig:heating_height} are from driver spectra
where the amplitudes have been changed so that the net Poynting flux is the same in all three cases. 
For the 10 Hz cut-off the total heating rate in the chromosphere is only 
of the order of $10^3 \mathrm{erg \:cm^{-2} \:s^{-1}}$. The majority of the input energy in these simulations goes into increasing the energy stored
in long wavelength Alfv\'en waves in the chromosphere and not into heating. A fraction of the total input power does leak into the corona, see later, 
but the continual buildup of Alfv\'en wave energy at long wavelength throughout the simulations, while correct for the incompressible simulations,
is unphysical and can only be corrected by the inclusion of compressibility. 

\begin{figure}
\includegraphics[scale=0.5]{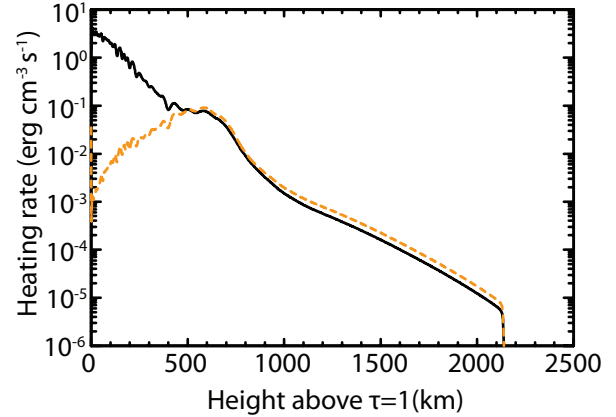}
\caption{Heating rate as a function of height for the full incompressible simulation including Pedersen resistivity $\eta_p$, equation \ref{Pedersen}, 
(black solid line) or Cowling resistivity only, $\eta_c=\eta_p-\eta$ (orange dashed line) with a 0.1Hz cut-off in the driver spectrum.}
\label{fig:resist_height}
\end{figure}

As with \citet{Goodman2011,Tu2013} it is found that heating in the photosphere and low chromosphere is dominated by dissipation due to 
classical resistivity 
(Figure~\ref{fig:resist_height}) while heating in the upper chromosphere is mainly due to Cowling resistivity. The largest heating rates 
occur in the parts of the simulation that are dominated by electron-ion 
and electron-neutral resistivity. 

The qualitative features of Figure~\ref{fig:heating_height} can be reproduced with a simple analytic
approach. This assumes that the heating is from a single pass of Alfv\'en waves
and all wavelengths are less than the density scale height. If only Alfv\'en waves are present then the Alfv\'en wave energy $U(y)=\rho(y) v_z^2(y)$
is determined by
\begin{equation}
	\frac{\partial U}{\partial t} + \frac{\partial S}{\partial y} 	= -\eta j^2
\end{equation}
with $S$ given by Equation~(\ref{eq:approxpoynt}) and $\eta$ the total resistivity. Since $S=c_A U$ in steady state, this can be solved to give
\begin{equation}
	U(y) = U(0) \frac{c_A(0)}{c_A(y)}\exp \Bigl[ -\omega^2 \int_0^y{\frac{\eta(y')}{\mu_0 c_A(y')^3}}\ dy'\Bigr].
\end{equation}
In this linearized WKB approach with only upward propagating Alfv\'en waves the local heating rate $h(y,\omega)$ is given by
\begin{equation}
	h(y,\omega)=U(0) \frac{\eta(y)c_A(0)}{\mu_0 c_A(y)^3}\exp(-\omega^2 F(y)),	
	\label{eq:heating_analytic}
\end{equation}
where
\begin{equation}
	F(y) = \int_0^y{\frac{\eta(y')}{\mu_0 c_A(y')^3}}\ dy'.
\end{equation}
Integrating over $\omega$ then gives the local heating rate for an 
incompressible chromosphere under the assumption that all the heating is from a single upward propagating pass through the chromosphere 
and transition region, i.e. ignoring reflection. This analytic heating rate is plotted in Figure~\ref{fig:heating_height}. The discrepancy between the 
simplified analytic model and full simulations are due to the WKB approximation and the absence of reflected waves in the analysis leading to 
Equation~(\ref{eq:heating_analytic}). 
Also plotted on figure \ref{fig:heating_height} is an
estimate of the required volumetric heating rate as a function of height
taken from \citep{Avrett1981}. This heating rate is not well defined below the temperature minimum.

\begin{figure}
\includegraphics[scale=0.5]{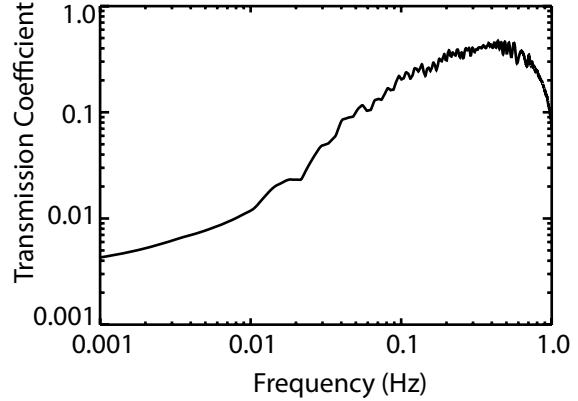}
\caption{Spectral energy transmission coefficient of the photospheric cavity
from incompressible simulations}
\label{trans_coeff}
\end{figure}

For the incompressible simulations there is no coupling between modes in 1.5D. As a result it is easy to estimate the effective transmission coefficient
of the transition region as a function of frequency without the complication which would arise from mode coupling in a non-linear simulation. 
The transmission coefficient is calculated as the ratio of the Poynting flux, as a function of frequency, just above the transition region to the 
Poynting flux at the base of the simulation domain. This is then smoothed with a boxcar moving average of 0.005 Hz width and plotted in 
Figure~\ref{trans_coeff}. Transmission ranges between 40\% at 0.5Hz to 0.32\% at 0.001Hz. Only higher frequencies have appreciable transmission
coefficients as the shorter wavelengths 'see' a density ramp rather than a discontinuity at the transition region. For frequencies above 0.5 Hz
the measured transmission drops due to these frequencies been damped by Pedersen resistivity before they reach the transition region. 

The overall energy budget from incompressible simulations is therefore that high frequency, i.e. greater than 0.1 Hz, Alfv\'en waves can both heat the
upper chromosphere and leak into the corona. The majority of the chromospheric heating in incompressible simulations come from these
high-frequency waves. For the chosen power law spectrum with r.m.s. amplitude $\sim 400 \mathrm{m \:s^{-1}}$ at the photosphere the typical heating rate 
above 1000 km is of the order of 
$10^{-6} - 10^{-5} \mathrm{erg \:cm^{-3}\:s^{-1}}$ for a 50 G vertical field. The peak chromospheric heating required to balance radiative and conductive
losses are $\sim 0.1 \mathrm{erg \:cm^{-3}\:s^{-1}}$ with rates of $\sim 10^{-3} \mathrm{erg \:cm^{-3}\:s^{-1}}$ needed in the upper chromosphere \citep{Ulmschneider1974}.
Higher heating rates could be achieved with the Kolmogorov driver used here by choosing a stronger background field strength or alternatively a higher
r.m.s. velocity driver could be used. Either of these would however increase the Poynting flux in the observed low-frequency part of the
spectrum and thus contradict observations.
Another possibility is that distribution of Alfv\'en waves deviates significantly from a $-5/3$ power law and has more energy 
at higher frequencies. This would require orders of magnitudes more energy in the high-frequencies contradicting our assumption of a turbulent
photospheric driver with a power-law spectra and is not studied further here.

\section{Compressible model}
\label{sec:compressible}
The incompressible model in 1.5D cannot be turbulent and there is no cascade of energy to smaller scales. The heating rate from Pedersen 
resistivity can be made to match the chromospheric heating requirements by ensuring sufficient energy is input into the base of the simulation in
frequencies which are damped by ion-neutral collisions. The heating rates as a function of height are therefore entirely dependent on the user-specified 
input spectrum which for these frequencies is poorly constrained by observations. Work by Hollweg \citep{Hollweg1982} and others have
also shown the importance of ponderomotive coupling of Alfv\'en waves to slow and fast modes in a compressive medium. In this section we repeat 
the simulations of Section~\ref{sec:incompressible} for the full set of compressible MHD equations. As before the heating due to Pedersen
resistivity is calculated but not added into the simulation thermal energy. This approach is not possible for the shock heating which is allowed 
in compressible simulation, as this is required to ensure the correct shock jump conditions. An inevitable consequence of this is that the model
atmosphere heats up due to shock dissipation. The Pedersen resistivity is however kept at the same profile as derived from the initial conditions.
In the absence of conductive and radiative losses the shock heating would ionize the chromosphere and turn-off Cowling resistivity, i.e.
that part of the Pedersen resistivity due to ion-neutral collisions. This
lack of self-consistency is shown below to not be significant as the dominant heating is through shock dissipation not resistivity.

\begin{figure}
\includegraphics[scale=0.5]{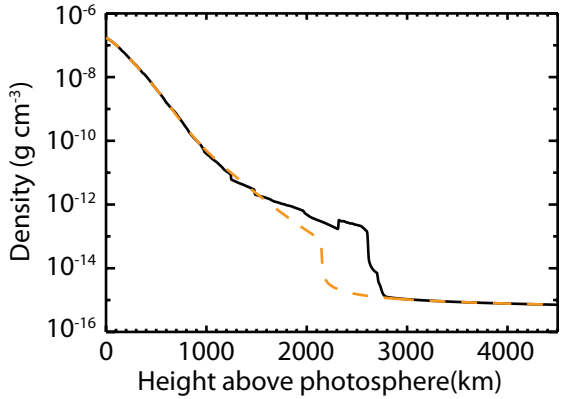}
\caption{Density at the start of the simulation (orange dashed line) and at 20 Alfv\'en transit times of the photospheric 
cavity (black line) from compressible simulations.}
\label{fig:density}
\end{figure}

\begin{figure}
\includegraphics[scale=0.5]{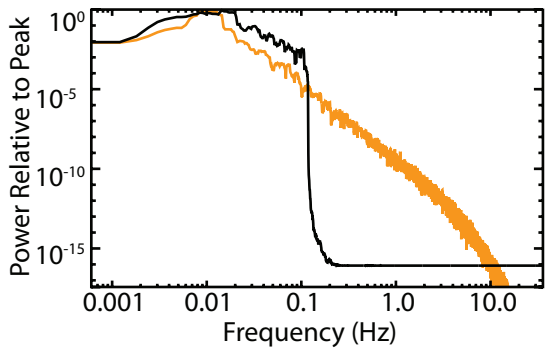}
\caption{Spectral energy in driver at the base of the simulation domain (black line) and in the upper chromosphere (orange line) from compressible simulations}. 
Both lines are smoothed with a boxcar moving average of width 0.05~Hz.
\label{comp_spect_en}
\end{figure}

A limitation of the 1.5D model used here is that the net ponderomotive force from a persistent photospheric driver lifts the 
transition region
and changes the atmospheric density structure. This is shown in Figure~\ref{fig:density}. 
In a compressible simulation energy cascades to shorter scale-lengths are now permitted. This can clearly be seen in 
Figure~\ref{comp_spect_en}, 
where the spectral energy in Alfv\'en waves from a driver introducing wave energy between 0.001~Hz and 0.1~Hz (black line) 
is compared to the 
energy present in the upper chromosphere (2000~km above the photospheric surface). Power at frequencies above the driver cutoff are clearly 
present in the upper chromosphere. This is not a purely
Alfv\'enic cascade as it shows the characteristic -2 spectral power of a shock dominated system, but it does provide a mechanism by 
which shorter length scales can be created self-consistently within the simulation. As a check of the simulations for consistence 
against observations we note that the velocity field 2000km above the photosphere in these simulations is $23\pm 2$ km/s. Transverse 
velocities observed for spicules by De Pontieu et al. \citet{DePontieu2007} 
were $20\pm 5$ km/s.

\begin{figure}
\includegraphics[scale=0.5]{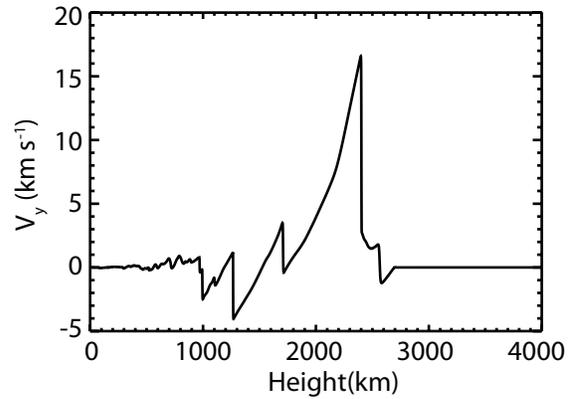}
\caption{Velocity component parallel to the background magnetic field with height, showing the formation of discontinuities from compressible simulations.}
\label{shock_plot}
\end{figure}

\begin{figure}
\includegraphics[scale=0.5]{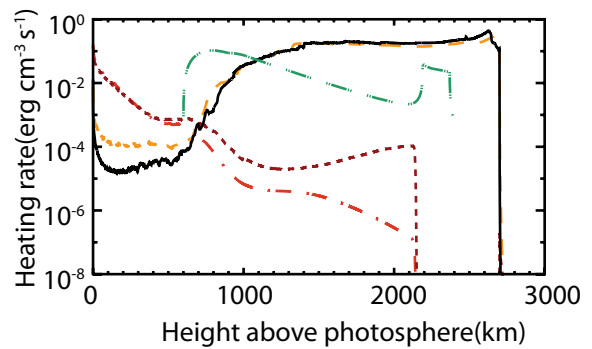}
\caption{Heating rate with height in compressible simulations of two different resolutions. The orange dashed line is shock heating 
in a simulation of 8192 grid cells with a driver cutoff frequency of 1~Hz, the black solid line shock heating in a simulation of 
16872 grid cells with the same cutoff frequency. The dark red short dashed line is the resistive heating from the 16872 cell 
compressible simulation, and the light red dash-dot line is from an equivalent incompressible simulation.
The green dot-dot-dash line is the estimated heating requirement
for the quite chromosphere taken from \citep{Avrett1981}.}
\label{heating_compress}
\end{figure}

Figure~\ref{shock_plot} shows that slow mode shocks are produced in the simulations. Since the entropy jump across the shock 
can heat the plasma, care must be taken to correctly reproduce this shock heating.  In order to ensure that shocks in the simulations 
satisfy the entropy condition in the shock jump relations, LARE uses the artificial viscosity formulation of \citet{Caramana1998,Arber2001} 
which both viscously changes the velocity and deposits the associated viscous heating. In the limit of infinite resolution this 
viscosity goes to zero in smooth regions of the solution and satisfies the shock jump conditions at discontinuities. However, at 
finite resolution viscosity is erroneously applied to parts of the solution which are steep on the scale of the grid, but are not 
true discontinuities. This means that careful convergence testing of the results of the simulations is required to confirm that the 
heating is due to shocks and not due to numerical error. This is shown in Figure~\ref{heating_compress} where it is shown that shock 
heating is more important in the upper chromosphere than resistive heating, and also that the shock heating rate is converged in these 
simulations to within 10\% in the chromosphere. While shock heating rates in the photosphere are not converged, they are clearly lower 
than the well-converged resistive heating rate at those heights. The light red dash dot line on Figure~\ref{heating_compress} shows the 
heating rate from the incompressible simulations. It can be seen that the overall resistive heating rate is similar in both compressible 
and incompressible simulations, but that additional resistive heating in the upper chromosphere is observed in the compressible 
simulations. Despite this, it is still 4 orders of magnitude lower than the shock heating rate.

\begin{figure}
\includegraphics[scale=0.5]{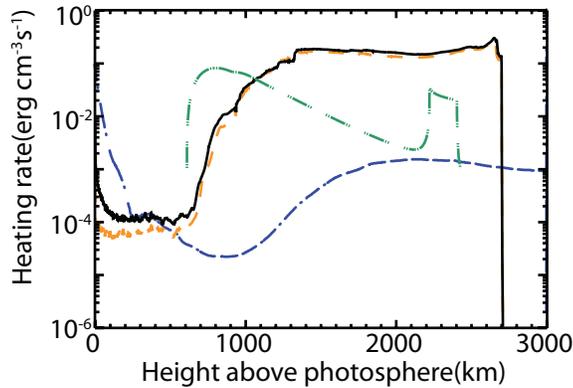}
\caption{Heating rate with height in compressible simulations with three driver upper cutoff frequencies. Black solid line has 
upper cutoff frequency of 0.1~Hz, orange dashed line of 1~Hz and blue line 0.01Hz. 
The green dot-dot-dash line is the estimated heating requirement
for the quite chromosphere taken from \citep{Avrett1981} based on observations.}
\label{visc_freq}
\end{figure}

Figure~\ref{visc_freq} shows that shock heating in the compressible simulations is the same for both cut-off frequencies above 0.1 Hz. No results are
shown for a cut-off of 10 Hz as these are the same as those for 1 Hz. The shock heating 
rate is higher than the heating rate needed to balance conduction and radiation in the chromosphere for a realistic driver amplitude. This may be explained by artificial effects of these 1.5D simulations. In particular, the missing vertical structure of the magnetic field and the 
artificial enhancement of shock strength due to shock coalescence - only possible in 1.5D. It is therefore possible to limit the heating due to 
Pedersen resistivity
by imposing a cut-off on photospheric driver frequencies of 0.1 Hz while maintaining the same level of shock heating. The extent to which this 
may be true
in a realistic chromosphere cannot be addressed in 1.5D models and must be tested in 2D and 3D. For a driver 
cut-off frequency of 0.01 Hz the shock heating is reduced by three orders of magnitude

\section{Conclusions}
\label{sec:conclusions}

The simulations presented in this paper show that it may be possible to heat the solar chromosphere by dissipating Alfv\'en wave energy. 
If energy is present in high enough frequency modes then resistive heating from Pedersen resistivity is sufficient to heat the solar 
chromosphere directly. The presence of such high frequency modes is not currently observable. However, if compressive effects are 
included then shock heating is capable of heating the chromosphere for a broad range of driver spectrum in a 1.5D model. In compressible 
simulations shock heating dominates over resistive dissipation in the upper chromosphere even for driver spectra with high frequency 
components. A problem with all 1.5D models is that shocks can coalesce and as a result the shock heating reported here is certainly an over-estimate. Also
absent from the current model is the flux expansion expected in chromospheric flux tubes. In addition, the possibility that the Alfv\'en turbulent cascade, 
which requires at least two spatial dimensions and would be fastest in 3D, may increase the relative importance of resistive dissipation by increasing the rate
at which long wavelength Alfv\'enic energy cascades to the resistive dissipation scale. Despite these omissions from the current 1.5D modelling there are
still a number of important conclusions which can be drawn and questions asked to guide future 2D and 3D work.
\begin{itemize}
	\item Evidence from over the last 30 years, e.g. references from \citet{Hollweg1981} onwards, have all shown that a low-frequency photospheric driver
	can excite Alfv\'en waves which are of sufficient energy to heat the chromosphere through shock heating. The slow modes are generated in the chromosphere
	via ponderomotive driving. The present study has focused on the chromosphere in more detail, and higher resolution, than previous studies, included
	Pedersen resistivity and confirmed this potential chromospheric heating mechanism. 
	\item While the incompressible results from \citet{Tu2013} are confirmed the observed heating as a function of height is entirely a function
	of the user specified driver since no energy cascade is present. Furthermore this is not changed by the inclusion of the Hall term. Most importantly,
	the results presented here show that compressible effects dominate the heating by orders of magnitude over Pedersen dissipation
	in the upper chromosphere.
	\item A turbulent cascade leading to heating \citep{Hollweg1986,Ballegooijen2011} is a possible heating mechanism where Alfv\'en wave energy can be thermalized. 
	However, compressible MHD effects must be
	included to check if in multi-dimensions the dominant energy loss from a low-frequency driver is through an Alfv\'en cascade or through coupling
	to slow modes. The present 1.5D study suggests that coupling to slow modes, which subsequently shock and heat the chromosphere, is the 
	dominant heating mechanism.
\end{itemize}  
In 2D and 3D the Alfv\'enic turbulent cascade to short scales, and hence dissipation, may be terminated at the dissipation scale of Pedersen
resistivity. Alternatively, Alfv\'enic wave energy may not reach these scales due to energy loss to slow modes via ponderomotive coupling, 
through mode coupling at the $\beta=1$ surface or geometric coupling through flux tube expansion.
While only 1.5D,
the current study strongly suggests ponderomotive coupling as the dominant mechanism by which Alfv\'en waves loose energy in the chromosphere. 
A final possibility
is that energy leaks through the transition region (Figure~\ref{comp_spect_en}) faster than it cascades to the dissipation scale. Only full simulations
in 2D and 3D will be able to determine which of these effects is most important.

\acknowledgments
This work was supported by STFC grant ST/L000733/1 and results were obtained using the DiRAC HPC Facility (www.dirac.ac.uk). 
This equipment was funded by BIS National E-infrastructure capital grant ST/K00042X/1, STFC capital grants ST/H008519/1 and 
ST/K00087X/1. Dr Shelyag is the recipient of an Australian Research Council's Future Fellowship (project number FT120100057). 
The authors are grateful for the financial support provided by Monash-Warwick Alliance Seed Fund.

\providecommand{\noopsort}[1]{}\providecommand{\singleletter}[1]{#1}%

\end{document}